\RequirePackage{lineno}
\documentclass[aps,prl,twocolumn,showpacs,superscriptaddress,floatfix]{revtex4}
\usepackage{epsfig}
\setpagewiselinenumbers

\newcommand {\snn}	{\sqrt{s_{_{\rm NN}}}}
\newcommand {\gev}	{{GeV/$c$}}
\newcommand {\nch}	{N_{\rm ch}}
\newcommand {\pp}	{{$p$+$p$}}
\newcommand {\pA}	{{$p$+A}}
\newcommand {\ppb}	{{$p$+Pb}}
\newcommand {\dAu}	{{$d$+Au}}

\newcommand {\FTPCd}	{FTPC-$d$}
\newcommand {\zyam}	{ZYAM}
\newcommand {\pt}	{p_{T}}

\newcommand {\dphi}	{\Delta\phi}
\newcommand {\deta}	{\Delta\eta}
\newcommand {\mean}[1]	{\langle #1\rangle}

\newcommand {\note}[1]	{}

\begin{document}
\title{Long-range pseudorapidity dihadron correlations in \dAu\ collisions at $\snn=200$~GeV}
\affiliation{AGH University of Science and Technology, Cracow 30-059, Poland}
\affiliation{Argonne National Laboratory, Argonne, Illinois 60439, USA}
\affiliation{Brookhaven National Laboratory, Upton, New York 11973, USA}
\affiliation{University of California, Berkeley, California 94720, USA}
\affiliation{University of California, Davis, California 95616, USA}
\affiliation{University of California, Los Angeles, California 90095, USA}
\affiliation{Universidade Estadual de Campinas, Sao Paulo 13131, Brazil}
\affiliation{Central China Normal University (HZNU), Wuhan 430079, China}
\affiliation{University of Illinois at Chicago, Chicago, Illinois 60607, USA}
\affiliation{Creighton University, Omaha, Nebraska 68178, USA}
\affiliation{Czech Technical University in Prague, FNSPE, Prague, 115 19, Czech Republic}
\affiliation{Nuclear Physics Institute AS CR, 250 68 \v{R}e\v{z}/Prague, Czech Republic}
\affiliation{Frankfurt Institute for Advanced Studies FIAS, Frankfurt 60438, Germany}
\affiliation{Institute of Physics, Bhubaneswar 751005, India}
\affiliation{Indian Institute of Technology, Mumbai 400076, India}
\affiliation{Indiana University, Bloomington, Indiana 47408, USA}
\affiliation{Alikhanov Institute for Theoretical and Experimental Physics, Moscow 117218, Russia}
\affiliation{University of Jammu, Jammu 180001, India}
\affiliation{Joint Institute for Nuclear Research, Dubna, 141 980, Russia}
\affiliation{Kent State University, Kent, Ohio 44242, USA}
\affiliation{University of Kentucky, Lexington, Kentucky, 40506-0055, USA}
\affiliation{Korea Institute of Science and Technology Information, Daejeon 305-701, Korea}
\affiliation{Institute of Modern Physics, Lanzhou 730000, China}
\affiliation{Lawrence Berkeley National Laboratory, Berkeley, California 94720, USA}
\affiliation{Massachusetts Institute of Technology, Cambridge, Massachusetts 02139-4307, USA}
\affiliation{Max-Planck-Institut fur Physik, Munich 80805, Germany}
\affiliation{Michigan State University, East Lansing, Michigan 48824, USA}
\affiliation{Moscow Engineering Physics Institute, Moscow 115409, Russia}
\affiliation{National Institute of Science Education and Research, Bhubaneswar 751005, India}
\affiliation{Ohio State University, Columbus, Ohio 43210, USA}
\affiliation{Institute of Nuclear Physics PAN, Cracow 31-342, Poland}
\affiliation{Panjab University, Chandigarh 160014, India}
\affiliation{Pennsylvania State University, University Park, Pennsylvania 16802, USA}
\affiliation{Institute of High Energy Physics, Protvino 142281, Russia}
\affiliation{Purdue University, West Lafayette, Indiana 47907, USA}
\affiliation{Pusan National University, Pusan 609735, Republic of Korea}
\affiliation{University of Rajasthan, Jaipur 302004, India}
\affiliation{Rice University, Houston, Texas 77251, USA}
\affiliation{University of Science and Technology of China, Hefei 230026, China}
\affiliation{Shandong University, Jinan, Shandong 250100, China}
\affiliation{Shanghai Institute of Applied Physics, Shanghai 201800, China}
\affiliation{Temple University, Philadelphia, Pennsylvania 19122, USA}
\affiliation{Texas A\&M University, College Station, Texas 77843, USA}
\affiliation{University of Texas, Austin, Texas 78712, USA}
\affiliation{University of Houston, Houston, Texas 77204, USA}
\affiliation{Tsinghua University, Beijing 100084, China}
\affiliation{United States Naval Academy, Annapolis, Maryland, 21402, USA}
\affiliation{Valparaiso University, Valparaiso, Indiana 46383, USA}
\affiliation{Variable Energy Cyclotron Centre, Kolkata 700064, India}
\affiliation{Warsaw University of Technology, Warsaw 00-661, Poland}
\affiliation{Wayne State University, Detroit, Michigan 48201, USA}
\affiliation{World Laboratory for Cosmology and Particle Physics (WLCAPP), Cairo 11571, Egypt}
\affiliation{Yale University, New Haven, Connecticut 06520, USA}
\affiliation{University of Zagreb, Zagreb, HR-10002, Croatia}

\author{L.~Adamczyk}\affiliation{AGH University of Science and Technology, Cracow 30-059, Poland}
\author{J.~K.~Adkins}\affiliation{University of Kentucky, Lexington, Kentucky, 40506-0055, USA}
\author{G.~Agakishiev}\affiliation{Joint Institute for Nuclear Research, Dubna, 141 980, Russia}
\author{M.~M.~Aggarwal}\affiliation{Panjab University, Chandigarh 160014, India}
\author{Z.~Ahammed}\affiliation{Variable Energy Cyclotron Centre, Kolkata 700064, India}
\author{I.~Alekseev}\affiliation{Alikhanov Institute for Theoretical and Experimental Physics, Moscow 117218, Russia}
\author{J.~Alford}\affiliation{Kent State University, Kent, Ohio 44242, USA}
\author{A.~Aparin}\affiliation{Joint Institute for Nuclear Research, Dubna, 141 980, Russia}
\author{D.~Arkhipkin}\affiliation{Brookhaven National Laboratory, Upton, New York 11973, USA}
\author{E.~C.~Aschenauer}\affiliation{Brookhaven National Laboratory, Upton, New York 11973, USA}
% ...

\collaboration{STAR Collaboration}\noaffiliation

\begin{abstract}
Dihadron angular correlations in \dAu\ collisions at $\snn=200$~GeV are reported as a function of the measured zero-degree calorimeter neutral energy and the forward charged hadron multiplicity in the Au-beam direction. A finite correlated yield is observed at large relative pseudorapidity ($\deta$) on the near side (i.e.~relative azimuth $\dphi\sim0$). This correlated yield as a function of $\deta$ appears to scale with the dominant, primarily jet-related, away-side ($\dphi\sim\pi$) yield. The Fourier coefficients of the $\dphi$ correlation, $V_n=\mean{\cos n\dphi}$, have a strong $\deta$ dependence. 
In addition, it is found that $V_1$ is approximately inversely proportional to the mid-rapidity event multiplicity, while $V_2$ is independent of it with similar magnitude in the forward ($d$-going) and backward (Au-going) directions.
\end{abstract}
\pacs{25.75.-q, 25.75.Dw}
\maketitle

Relativistic heavy-ion collisions are used to study quantum chromodynamics (QCD) at high energy densities at the Relativistic Heavy Ion Collider (RHIC) and the Large Hadron Collider (LHC)~\cite{Arsene:2004fa,Back:2004je,Adams:2005dq,Adcox:2004mh,Muller:2012zq}. 
Final-state particle emission in such collisions is anisotropic, quantitatively consistent with hydrodynamic flow resulting from the initial-state overlap geometry~\cite{Ollitrault:1992bk,Heinz:2013th}. Two-particle correlations are widely used to measure anisotropic flow and jet-like correlations~\cite{Wang:2013qca}. 
A near-side long-range correlation (at small relative azimuth $\dphi$ and large relative pseudorapidity $\deta$), called the ``ridge,'' has been observed after elliptic flow subtraction in central heavy-ion collisions at RHIC and the LHC~\cite{Adams:2005ph,Abelev:2009af,Alver:2009id,Abelev:2009jv,Chatrchyan:2013nka,ALICE:2011ab}. It is attributed primarily to triangular flow, resulting from a hydrodynamic response to initial geometry fluctuations~\cite{Alver:2010gr,Adamczyk:2013waa}.

As reference, \pp, \pA\ and \dAu\ collisions are often used to compare with heavy-ion collisions. 
Hydrodynamics is not expected to describe these small-system collisions.
However, a large $\deta$ ridge has been observed in high-multiplicity \pp~\cite{Khachatryan:2010gv} and \ppb~\cite{CMS:2012qk,Abelev:2012ola,Aad:2012gla,ABELEV:2013wsa} collisions at the LHC after a uniform background subtraction. The similarity to the heavy-ion ridge is suggestive of a hydrodynamic description of its origin, in conflict with early expectations. Indeed, hydrodynamic calculations with event-by-event fluctuations can describe the observed ridge and attribute it to elliptic flow~\cite{Bozek:2010pb,Bozek:2012gr}. 
Other physics mechanisms are also possible, such as the color glass condensate where the two-gluon density is enhanced at small $\dphi$ over a wide range of $\deta$~\cite{Dumitru:2010iy,Dusling:2012wy,Dusling:2013oia}, or quantum initial anisotropy~\cite{Molnar:2014mwa}. 

Furthermore, a back-to-back ridge is revealed by subtracting dihadron correlations in low-multiplicity \ppb\ from those in high-multiplicity collisions at the LHC~\cite{Abelev:2012ola,Aad:2012gla,ABELEV:2013wsa}. A similar double ridge is observed in \dAu\ collisions at RHIC by PHENIX within $0.48<|\deta|<0.70$ using the same subtraction technique~\cite{Adare:2013piz}. A recent STAR analysis has challenged the assumption of this subtraction procedure that jet-like correlations are equal in high- and low-multiplicity events~\cite{Adamczyk:2014fcx}. 
It was shown that the double ridge at these small-to-moderate $\deta$ has a significant contribution from residual jet-like correlations despite performing event selections via
forward multiplicities~\cite{Adamczyk:2014fcx}. A recent PHENIX study of large $\deta$ correlations, without relying on the subtraction technique, suggests a long-range correlation consistent with hydrodynamic anisotropic flow~\cite{Adare:2014keg}. 
In order to further understand the underlying physics mechanism, here in this Letter, we present our results on long-range (large $\deta$) correlations in \dAu\ collisions at $\snn=200$~GeV as a function of $\deta$ and the event multiplicity.
The large acceptance of the STAR detector is particularly well suited for such an analysis over a wider range in $\deta$.

The data were taken during the \dAu\ run in 2003 by the STAR experiment~\cite{Adams:2003im,Abelev:2008ab}. The details of the STAR detector can be found in Ref.~\cite{Ackermann:2002ad}. 
Minimum-bias \dAu\ events were triggered by coincidence of signals from the Zero Degree Calorimeters (ZDC)~\cite{Adler:2003sp} and the Beam-Beam Counters (BBC)~\cite{Ackermann:2002ad}. Particle tracks were reconstructed in the Time Projection Chamber (TPC)~\cite{Anderson:2003ur} and the forward TPC (FTPC)~\cite{Ackermann:2002yx}. The primary vertex was determined from reconstructed tracks. In this analysis, events were required to have a primary vertex position $|z_{\rm vtx}|<50$~cm from the TPC center along the beam axis. 
TPC(FTPC) tracks were required to have at least 25(5) out of the maximum possible 45(10) hits and a distance of closest approach to the primary vertex within 3~cm. 
\note{The TPC and FTPC track momentum resolutions were .... $0.0078 + 0.0098 \cdot \pt$ (\gev)~\cite{Anderson:2003ur}.}

Three measurements were used to select \dAu\ events: neutral energy 
by the ZDC and charged particle multiplicity within $-3.8<\eta<-2.8$ by the FTPC~\cite{Adams:2003im,Abelev:2008ab}, both in the Au-beam direction, and charged particle multiplicity within $|\eta|<1$ by TPC. 
Weak but positive correlations were observed between these measurements; the same event fraction defined by these measures corresponded to significantly different \dAu\ event samples~\cite{Adamczyk:2014fcx}. 
In this work we study 0-20\% high-activity and 40-100\% low-activity collisions according to each measure.


The pairs of particles used in dihadron correlations are customarily called the trigger and the associated particle. Two sets of dihadron correlations are analyzed: TPC-TPC correlations where both the trigger and associated particles are from the TPC ($|\eta|<1$), and TPC-FTPC correlations where the trigger particle is from the TPC but the associated particle is from either the FTPC-Au ($-3.8<\eta<-2.8$) or \FTPCd\ ($2.8<\eta<3.8$). The $\pt$ ranges of the trigger and associated particles are both $1<\pt<3$~\gev. The associated particle yields are normalized per trigger particle. The yields are corrected for the TPC and FTPC associated particle tracking efficiencies of $85\%\pm5\%$~(syst.)~and $70\%\pm5\%$~(syst.), respectively, 
which do not depend on the event activity in \dAu\ collisions~\cite{Abelev:2008ab,Adams:2003im}.

The detector non-uniformity in $\dphi$ is corrected by the event-mixing technique, where a trigger particle from one event is paired with associated particles from another event. The mixed events are required to be within 1~cm in $z_{\rm vtx}$, with the same multiplicity (by FTPC-Au or TPC) or 
similar energy (by ZDC-Au). The mixed-event correlations are normalized to 100\% at $\deta=0$ for TPC, and at $\pm3.3$ for \FTPCd\ and FTPC-Au associated particles, respectively.

Two analysis approaches are taken. One is to analyze the correlated yields after subtracting a uniform combinatorial background. The background normalization is estimated by the Zero-Yield-At-Minimum (\zyam) assumption~\cite{Adams:2005ph,Ajitanand:2005jj}. \zyam\ is taken as the lowest yield averaged over a $\dphi$ window of $\pi/8$ radian width, after the correlated yield distribution is folded into the range of $0<\dphi<\pi$. 
The \zyam\ systematic uncertainty is estimated by the yields averaged over windows of half and three half the width. 
We also fit the $\dphi$ correlations by two Gaussians (with centroids fixed at 0 and $\pi$) plus a pedestal. The fitted pedestal is consistent with \zyam\ within the statistical and systematic errors because the near- and away-side peaks are well separated in \dAu\ collisions.
The systematic uncertainties on the correlated yields are taken as the quadratic sum of the \zyam\ and tracking efficiency systematic uncertainties. 
The other approach is to analyze the Fourier coefficients of the $\dphi$ correlation functions, $V_n=\mean{\cos n\dphi}$. No background subtraction is required. Systematic uncertainties on the Fourier coefficients 
are estimated, by varying analysis cuts, to be less than 10\% \note{5\%} for $V_1$ and $V_2$, and smaller than the statistical errors for $V_3$. 

\begin{figure*}
  \begin{center}
    \includegraphics[width=0.9\textwidth]{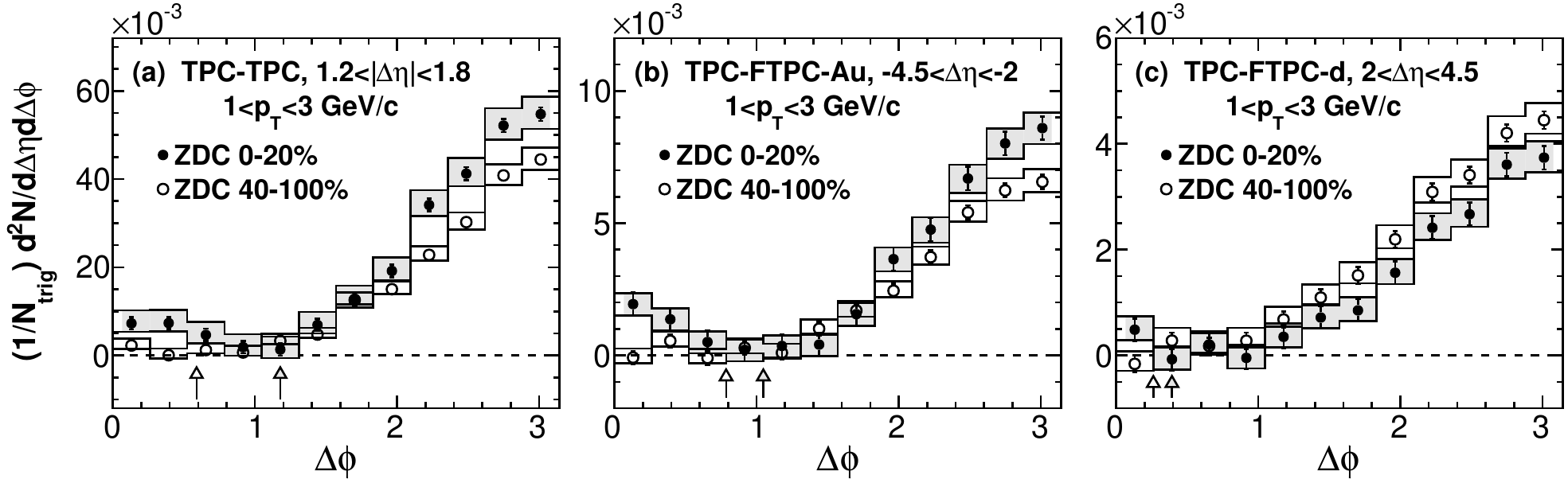}
  \end{center}
  \caption{Correlated dihadron yield, per radian per unit of pseudorapidity, as a function of $\dphi$ in three ranges of $\deta$ in \dAu\ collisions. Shown are both low and high ZDC-Au activity data. Both the trigger and associated particles have $1<\pt<3$~\gev. The arrows indicate \zyam\ normalization positions. The error bars are statistical and histograms indicate the systematic uncertainties.}
  \label{fig:dphi}
\end{figure*}

Figure~\ref{fig:dphi} shows the \zyam-subtracted correlated yields as a function of $\dphi$ in ZDC-Au low- and high-activity \dAu\ collisions. 
The TPC-TPC correlation at large $\deta$ is shown in panel (a), whereas the TPC-FTPC correlations are shown in panels (b) and (c) for Au- and $d$-going directions, respectively. The \zyam\ statistical error is included as part of the systematic uncertainty drawn in Fig.~\ref{fig:dphi} because it is common to all $\dphi$ bins. No difference is observed in TPC-TPC correlations between positive and negative $\deta$, so they are combined in Fig~\ref{fig:dphi}(a). The away-side correlated yields are found to be larger in high- than low-activity \dAu\ collisions for TPC and FTPC-Au correlations. The opposite behavior is observed for the \FTPCd\ correlations, Fig.~\ref{fig:dphi}(c). 

On the near side, the correlated yields are consistent with zero in the low-activity events and, in \FTPCd, in the high-activity events as well. (Note that the yield value cannot be negative because of the \zyam\ assumption.)
In contrast, in TPC and FTPC-Au, finite correlated yields are observed in high-activity events. A similar result was observed by PHENIX~\cite{Adare:2014keg}. 
In Fig.~\ref{fig:dphi}, the event activity is determined by ZDC-Au. For event activity determined by FTPC-Au or TPC multiplicity, the data are qualitatively similar. In Table~\ref{tab:yields}, the correlated yields integrated over the near side ($|\dphi|<\pi/3$) and the away side ($|\dphi-\pi|<\pi/3$), normalized by the integration range, are tabulated together with the \zyam\ magnitude for low- and high-activity events determined by the various measures.
\begin{table*}
  \caption{Near- ($|\dphi|<\pi/3$) and away-side ($|\dphi-\pi|<\pi/3$) correlated yields and \zyam\ background magnitude, per radian per unit of pseudorapidity, at large $\deta$ in low- and high-activity \dAu\ collisions. Positive(negative) $\eta$ corresponds to $d$(Au)-going direction. Both the trigger and associated particles have $1<\pt<3$~\gev. All numbers have been multiplied by $10^4$. Errors are statistical except the second error of each \zyam\ value which is systematic and applies also to the corresponding near- and away-side yields. An additional 5\% efficiency uncertainty applies.}
  \label{tab:yields}
  \begin{tabular}{r|lccc|lcccccc}\hline
    Event    & Event     &\multicolumn{3}{c|}{$1.2<|\deta|<1.8$} & Event     & \multicolumn{3}{c}{$-4.5<\deta<-2$} & \multicolumn{3}{c}{$2<\deta<4.5$} \\
    activity & selection & \zyam & near & away                   & selection & \zyam & near & away                 & \zyam & near & away \\\hline
    40-100\% & ZDC  & 1896$\pm$7$^{+1}_{-13}$ & 10$\pm$4 & 346$\pm$5 & ZDC & 978$\pm$2$^{+1}_{-2}$ & 2$\pm$1 & 55$\pm$1 & 361$\pm$1$^{+1}_{-2}$ & 1$\pm$1 & 38$\pm$1 \\
    0-20\%   &      & 3043$\pm$11$^{+15}_{-26}$ & 53$\pm$7 & 456$\pm$7 & & 1776$\pm$4$^{+2}_{-1}$ & 10$\pm$2 & 70$\pm$2 & 438$\pm$2$^{+1}_{-2}$ & 1$\pm$1 & 31$\pm$1 \\\hline
    40-100\% & FTPC & 1324$\pm$7$^{+2}_{-6}$ & 7$\pm$4 & 347$\pm$5 & TPC & 636$\pm$2$^{+1}_{-2}$ & 6$\pm$1 & 59$\pm$1 & 309$\pm$2$^{+1}_{-1}$ & 3$\pm$1 & 45$\pm$1 \\
    0-20\%   &      & 3468$\pm$10$^{+7}_{-5}$ & 43$\pm$6 & 429$\pm$7 & & 1899$\pm$3$^{+2}_{-5}$ & 15$\pm$2 & 75$\pm$2 & 445$\pm$1$^{+1}_{-3}$ & 2$\pm$1 & 27$\pm$1 \\\hline
\end{tabular}
\end{table*}

For trigger particles in our $\pt$ range of $1<\pt<3$~\gev, the away-side correlation in \dAu\ collisions is expected to be dominated by jet-like correlations~\cite{Gyulassy:1994ew}. 
Inspecting the near-side correlation amplitude at large $\deta$, any possible non-jet, e.g.~anisotropic flow, contributions on the away side should be order of magnitude smaller.
Perhaps the observed away-side dependence on ZDC-Au event activity arises from a correlation between jet production and the forward beam remnants. Or, the underlying physics may be more complex; for example the opposite away-side trends in the Au- and $d$-going directions may arise from different underlying parton distributions in high- and low-activity collisions.
The finite correlated yield on the near side is, on the other hand, rather surprising because jet-like contributions should be minimal at these large $\deta$ distances. Hijing simulation~\cite{Gyulassy:1994ew} of \dAu\ collisions indicates that jet correlations within our $\pt$ range after \zyam\ background subtraction is consistent with zero at $|\deta|>1.5$.

\begin{figure}
  \begin{center}
    \includegraphics[width=0.4\textwidth]{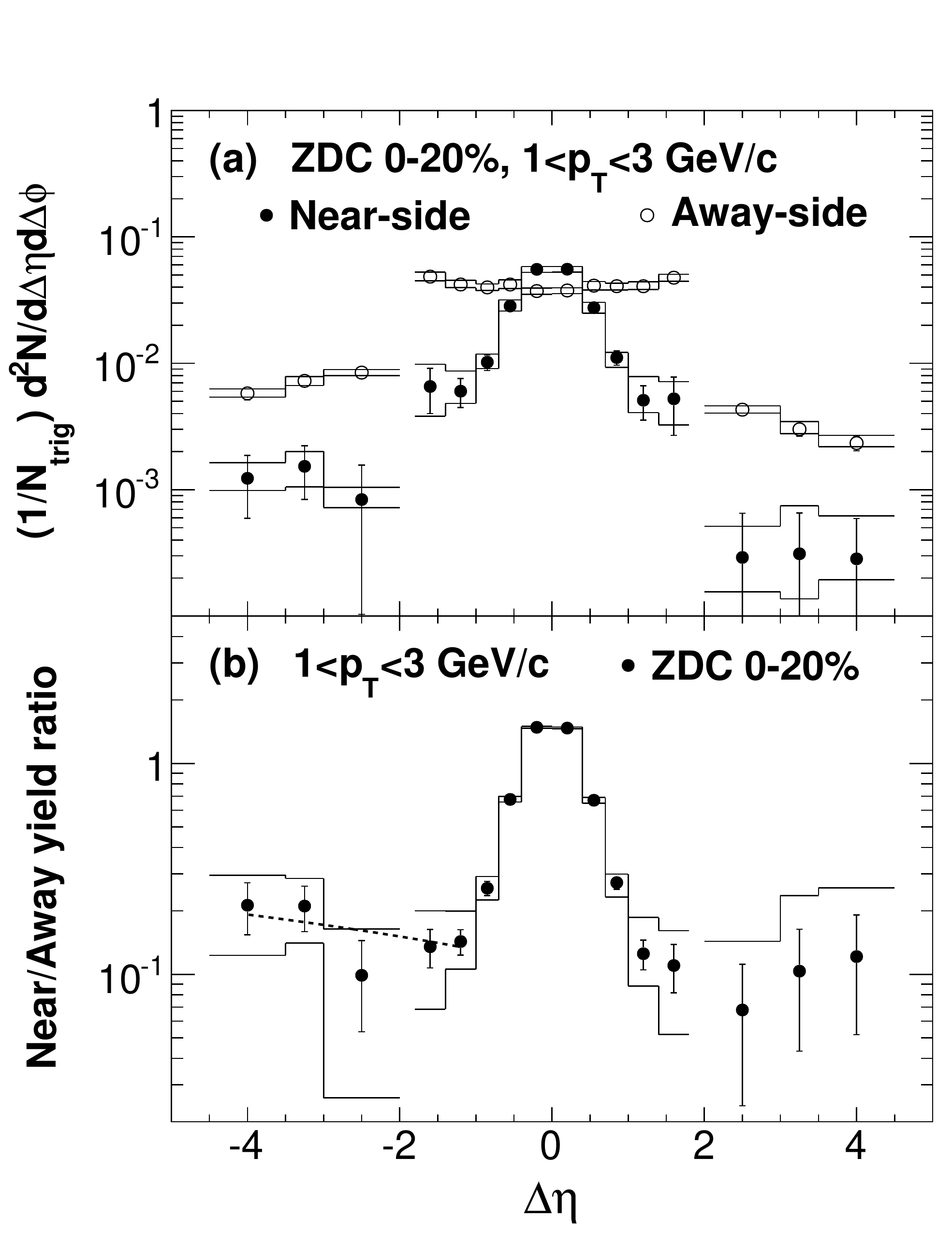}
  \end{center}
  \caption{The $\deta$ dependence of (a) the near- ($|\dphi|<\pi/3$) and away-side ($|\dphi-\pi|<\pi/3$) correlated yields, and 
    (b) the ratio of the near- to away-side correlated yields in \dAu\ collisions. 
    Positive(negative) $\eta$ corresponds to $d$(Au)-going direction. Only high ZDC-Au activity data are shown. 
    The error bars are statistical and histograms indicate the systematic uncertainties (for $\deta>2$ in (b) the lower bound falls outside the plot). 
    The dashed curve in (b) is a linear fit to the $\deta<-1$ data points.}
  \label{fig:deta}
\end{figure}

To study the $\deta$ dependence of the correlated yields in the TPC and FTPC, the correlation data are divided into multiple $\deta$ bins. In Fig.~\ref{fig:deta}(a), the near- and away-side correlated yields are shown as a function of $\deta$. To avoid auto-correlations, we have used ZDC-Au for event selections for both the TPC and FTPC correlation data. 
Unlike in Fig.~\ref{fig:dphi}, the \zyam\ statistical errors are dependent of $\deta$ and are therefore included in the statistical error bars of the data points. The away-side correlation shape, noticeably concaved for TPC, is presumably determined by the underlying parton-parton scattering kinematics.
On the near side, finite correlated yields are observed at large $\deta$ on the Au-going side in all bins, while the yields are consistent with zero on the $d$-going side. As aforementioned, similar results have been previously observed in heavy-ion~\cite{Adams:2005ph,Abelev:2009af,Alver:2009id,Abelev:2009jv,Chatrchyan:2013nka,ALICE:2011ab}, \pp~\cite{Khachatryan:2010gv}, and p+Pb collisions~\cite{CMS:2012qk,Abelev:2012ola,Aad:2012gla,ABELEV:2013wsa}. There, the trigger and associated particles were taken from the same $\eta$ region. As a result, the correlated yields were approximately uniform in $\deta$~\cite{Xu:2013sua}, and were dubbed the ``ridge.'' In the three groups of correlation data in Fig.~\ref{fig:deta}(a), the trigger particles come from the TPC, but the associated particles come from different $\eta$ regions. Significant differences in pair kinematics result in the steps at $\deta=\pm2$ even though their $\deta$ gaps are similar.
Despite this, for simplicity, we refer to the large $\deta$ correlated yields in our data also as the ``ridge.''

In order to elucidate the formation mechanism of the ridge, we study in Fig.~\ref{fig:deta}(b) the ratio of the near- to away-side correlated yields. 
Because the \zyam\ value is common for the near and away side, its statistical error is included as part of the systematic uncertainty; this part of the systematic uncertainty is uncorrelated between $\deta$ bins. While the large peak at $\deta\sim0$ is due to the near-side jet, the ratio at $\deta<-1$ is rather insensitive to $\deta$, whether the correlations are from TPC or FTPC-Au. A linear fit (dashed-line in Fig.~\ref{fig:deta}(b)) to those data points at $\deta<-1$ yields a slope parameter of 
$-0.023\pm0.019^{+0.020}_{-0.010}$ 
with $\chi^2/{\rm ndf}=2.6/3$, indicating that the ratio is consistent with a constant within one standard deviation. The rather constant ratio is remarkable, given the nearly order of magnitude difference in the away-side jet-like correlated yields across $\deta=-2$ due to the vastly different pair kinematics. Since the away-side correlated yields are dominated by jets~\cite{Gyulassy:1994ew}, 
the finite, $\deta$-independent ratio at $\deta<-1$ may suggest a connection between the near-side ridge and jet production, even though any possible jet contribution to the near-side ridge at $|\deta|>1$ should be minimal. On the other hand, the near-side ridge does not seem to scale with the \zyam\ value, which represents the underlying background. A linear fit to the ratio of the near-side correlated yield over \zyam\ in the same $\deta<-1$ region gives a slope parameter of 
$6.5\pm1.6^{+3.7}_{-2.1}\times10^{-3}$, significantly deviating from zero. 

\begin{figure}
  \begin{center}
    \includegraphics[width=0.4\textwidth]{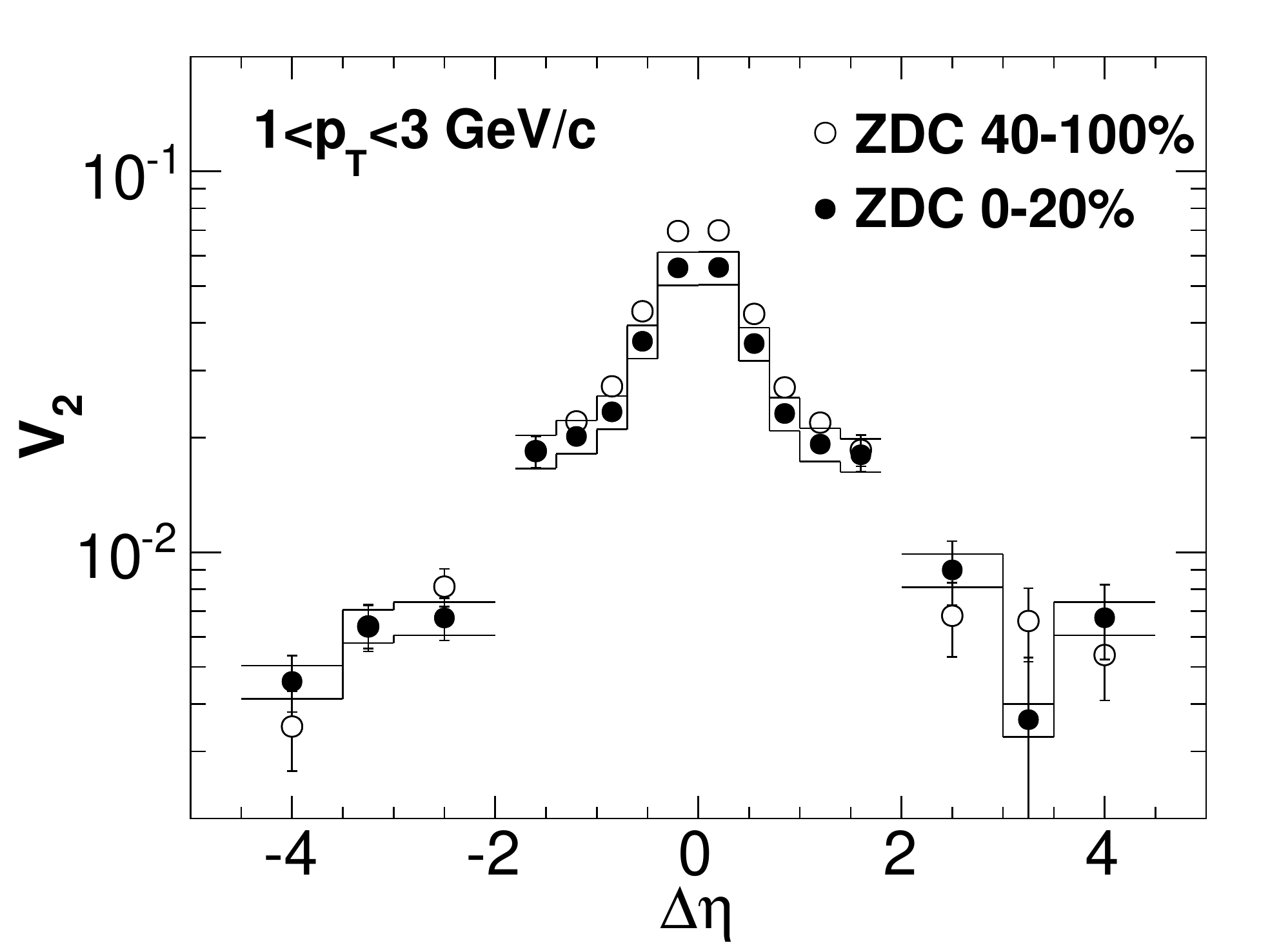}
  \end{center}
  \caption{The $\deta$ dependence of the second harmonic Fourier coefficient, $V_{2}$, in low and high ZDC-Au activity \dAu\ collisions. The error bars are statistical. Systematic uncertainties are 10\%\note{5\%} and are shown by the histograms, for clarity, only for the high-activity data.}
  \label{fig:detaVn}
\end{figure}

The correlated yields discussed above are subject to the \zyam\ background subtraction. Another way to quantify the ridge is via Fourier coefficients 
of the azimuthal correlation functions without background subtraction. 
Figure~\ref{fig:detaVn} shows the second harmonic Fourier coefficient ($V_2$) as a function of $\deta$ for both high and low ZDC-Au energy collisions. The $V_2$ values are approximately the same in high- and low-activity collisions at large $\deta$. Both decrease with increasing $|\deta|$ from the small $\deta$, jet dominated, region to the large $\deta$, ridge, region by nearly one order of magnitude.
The $\deta$ behavior of $V_2$, a measure of modulation relative to the average, is qualitatively consistent with the $\deta$-dependent ratio of the near-side correlated yield over \zyam. 
One motivation to analyze correlation data using Fourier coefficients is their independence of a ZYAM subtraction procedure. One way for $V_2$ to develop is through final-state interactions which, if prevalent enough, may be described in terms of hydrodynamic flow. 
If $V_2$ is strictly of a hydrodynamic elliptic flow origin, the data would imply a decreasing collective effect at backward/forward rapidities that is somehow independent of the activity level of the events. 

\begin{figure}
  \begin{center}
    \includegraphics[width=0.4\textwidth]{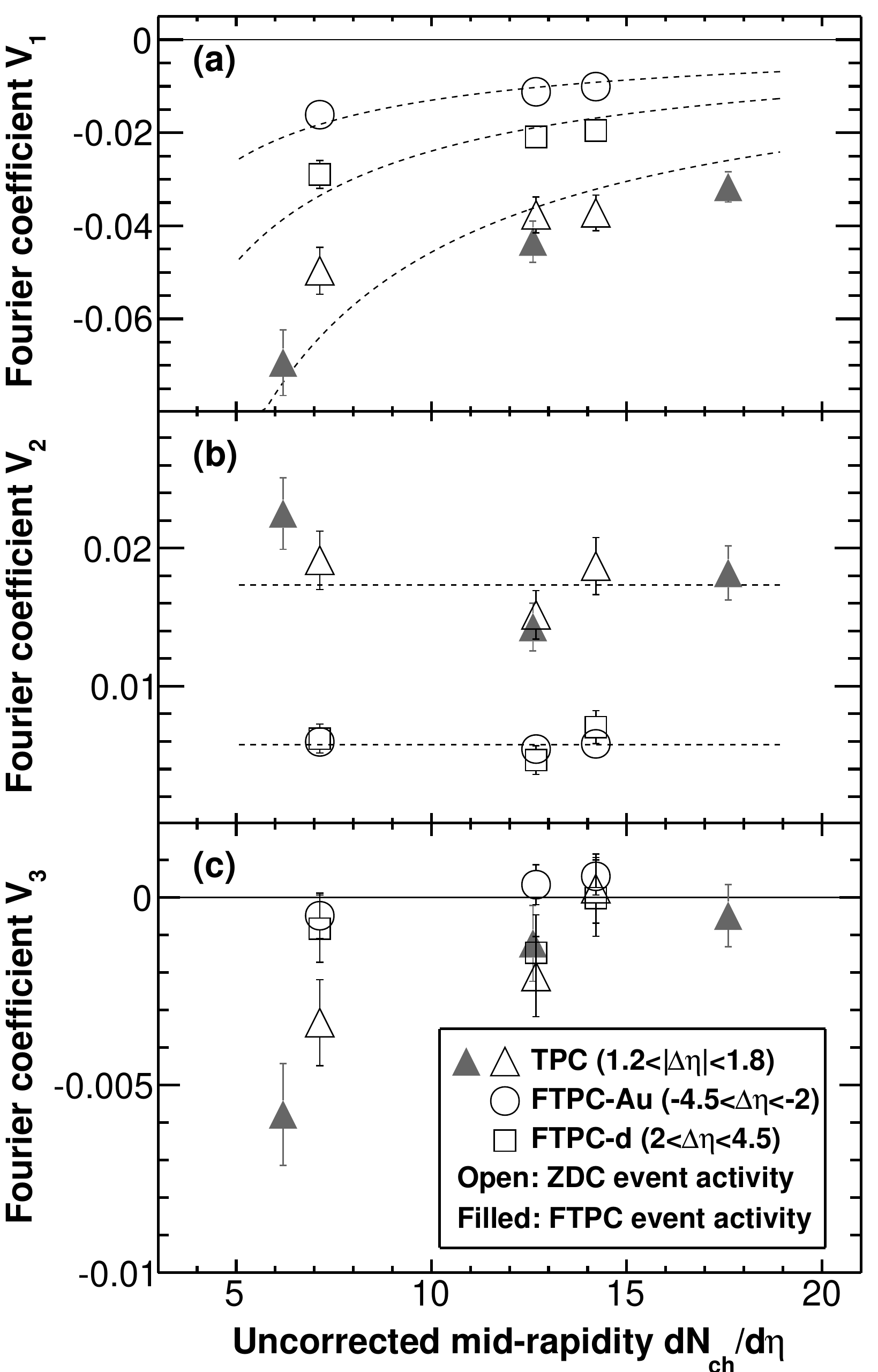}
  \end{center}
  \caption{Fourier coefficients (a) $V_1$, (b) $V_2$, and (c) $V_3$ versus the measured mid-rapidity charged particle $d\nch/d\eta$. Event activity selections by both ZDC-Au and FTPC-Au are shown. Trigger particles are from TPC, and associated particles from TPC (triangles), FTPC-Au (circles), and \FTPCd\ (squares), respectively. Systematic uncertainties are estimated to be 10\%\note{5\%} on $V_1$ and $V_2$, and smaller than statistical errors for $V_3$. Errors shown are the quadratic sum of statistical and systematic errors. The dashed curves are to guide the eye.}
  \label{fig:Vn}
\end{figure}

To gain further insights, the multiplicity dependencies of the first, second and third Fourier coefficients $V_1$, $V_2$ and $V_3$ are shown in Fig.~\ref{fig:Vn}. 
Three $\deta$ ranges are presented for FTPC-Au, TPC, and \FTPCd\ correlations, respectively. Results by both the ZDC-Au and FTPC-Au event selections are shown, plotted as a function of the corresponding measured charged particle pseudorapidity density at mid-rapidity 
$d\nch/d\eta$. 
The absolute value of the $V_1$ parameter in each $\deta$ range varies approximately as $(d\nch/d\eta)^{-1}$ (see the superimposed curves). This is consistent with jet contributions and/or global statistical momentum conservation. 
On the other hand, the $V_2$ parameter in each $\deta$ range is approximately independent of $d\nch/d\eta$ over the entire measured range (the dashed lines are to guide the eye).
Similar behavior of $V_2$ is also observed in \ppb\ collisions at the LHC~\cite{Aad:2013fja,Chatrchyan:2013nka,Abelev:2014mda}. 
Figure~\ref{fig:Vn} shows that the $V_3$ values are small and mostly consistent with zero, except for TPC-TPC correlation at the lowest multiplicity.

In \dAu\ collisions, dihadron correlations are dominated by jets, even at large $\deta$, where the away-side jet contributes~\cite{Gyulassy:1994ew}. The behavior of $V_1$ suggests that the jet contribution to $V_n$ is diluted by the multiplicity. The similar $V_2$ values and $\deta$ dependencies in different multiplicity collisions are, therefore, rather surprising. In order to accommodate a hydrodynamic contribution, there must be a coincidental compensation of the reduced jet contribution with increasing multiplicity, over the entire measured multiplicity range, by an emerging, non-jet contribution, such as elliptic flow. 

Whether or not a finite correlated yield appears on the near side depends on the interplay between $V_1$ and $V_2$ (higher order terms are negligible). Although the $V_2$ parameters are similar, the significantly more negative $V_1$ in low- versus high-multiplicity events eliminates the near-side $V_2$ peak in $\dphi$. The same applies also to the TPC-FTPC correlation comparison between the Au- and $d$-going directions. The $V_2$ values are rather similar for \FTPCd\ (forward rapidity) and FTPC-Au (backward rapidity) correlations, but the more negative $V_1$ for $d$-going direction eliminates the near-side $V_2$ peak. If the relevant physics in \dAu\ collisions is governed by hydrodynamics, then it may not carry significance whether or not there exists a finite near-side long-range correlated yield, which would be a simple manifestation of the relative $V_1$ and $V_2$ strengths.

Our $V_2$ data are qualitatively consistent with that from PHENIX~\cite{Adare:2014keg}. While PHENIX focused on the $\pt$ dependence, we study the Fourier coefficients as a function of $\deta$ afforded by the large STAR acceptance, as well as the event multiplicity. 
Hydrodynamic effects, if they exist in \dAu\ collisions, should naively differ over the measured multiplicity range and between Au- and $d$-going directions. However, the $V_2$ parameters are approximately constant over multiplicity, and quantitatively similar between the Au- and $d$-going directions. On the other hand, the correlation comparisons between low- and high-activity data reveal different trends for the Au- and $d$-going directions. The high- and low-activity difference in the FTPC-Au correlation in Fig.~\ref{fig:dphi}(b) may resemble elliptic flow, but that in the \FTPCd\ correlation in Fig.~\ref{fig:dphi}(c) is far from an elliptic flow shape. In combination, these data suggest that the finite values of $V_n$ cannot be exclusively explained by hydrodynamic anisotropic flow in \dAu\ collisions at RHIC.

In summary, dihadron angular correlations are reported for \dAu\ collisions at $\snn=200$~GeV as a function of the event activity from the STAR experiment. 
The event activity is classified by the measured zero-degree neutral energy in ZDC, the charged hadron multiplicity in FTPC, both in the Au-going direction, or the multiplicity in TPC. 
In a recent paper we have shown that the short-range jet-like correlated yield increases with the event activity~\cite{Adamczyk:2014fcx}. In this paper we focus on long-range correlations at large $|\deta|$, where jet-like contributions are minimal on the near side, although the away side is still dominated by jet production. Two approaches are taken, one to extract the correlated yields above a uniform background estimated by the \zyam\ method, and the other to calculate the Fourier coefficients, $V_n=\mean{\cos n\dphi}$, of the dihadron $\dphi$ correlations. 
The following points are observed:
(i) The away-side correlated yields are larger in high- than in low-activity collisions in the TPC and FTPC-Au, but lower in \FTPCd; 
(ii) Finite near-side correlated yields are observed at large $\deta$ above the estimated \zyam\ background in high-activity collisions in both the TPC and FTPC-Au (referred to as the ``ridge''); 
(iii) The ridge yield appears to scale with the away-side correlated yield at the corresponding $\deta<-1$, which is dominated by the away-side jet; 
(iv) The $V_2$ coefficient decreases with increasing $|\deta|$, but remains finite at both forward and backward rapidities ($|\deta|\approx3$) with similar magnitude; 
(v) The $V_1$ coefficient is approximately inversely proportional to the event multiplicity, but the $V_2$ appears to be independent of it. 
While hydrodynamic elliptic flow is not excluded with a coincidental compensation of jet dilution by increasing flow contribution with multiplicity and an unexpected equality of elliptic flow between forward and backward rapidities, the data suggest that there exists a long-range pair-wise correlation in \dAu\ collisions that is correlated with dijet production.

We thank the RHIC Operations Group and RCF at BNL, the NERSC Center at LBNL and the Open Science Grid consortium for providing resources and support. This work was supported in part by the Offices of NP and HEP within the U.S.~DOE Office of Science, the U.S.~NSF, the Sloan Foundation, the DFG cluster of excellence `Origin and Structure of the Universe' of Germany, CNRS/IN2P3, STFC and EPSRC of the United Kingdom, FAPESP CNPq of Brazil, Ministry of Ed.~and Sci.~of the Russian Federation, NNSFC, CAS, MoST, and MoE of China, GA and MSMT of the Czech Republic, FOM and NWO of the Netherlands, DAE, DST, and CSIR of India, Polish Ministry of Sci.~and Higher Ed., Korea Research Foundation, Ministry of Sci., Ed.~and Sports of the Rep.~Of Croatia, Russian Ministry of Sci.~and Tech, and RosAtom of Russia.

\bibliography{ref_ridge_20150224}
\end{document}